# C$_5$ as simplest ultrahard allotrope with mixed sp$^2$/sp$^3$ carbon hybridizations from first principles.

Samir F. Matar[*]

Lebanese German University (LGU), Sahel-Alma, Jounieh (P.O.Box 206), Lebanon.
https://orcid.org/0000-0001-5419-358X

[*] Email: s.matar@lgu.edu.lb

**Abstract**

*From crystal chemistry rationale and density functional DFT calculations, novel tetragonal carbon C$_5$ is proposed as simplest ultrahard allotrope with mixed hybrid carbon hybridizations (sp2/sp3). Novel pentacarbon is identified as cohesive and stable both dynamically and mechanically. Whereas charge density is localized about tetrahedral C-sp$^3$, it is found delocalized around trigonal C-sp$^2$ resulting in metallic behavior. The anisotropic structure characteristics caused by the aligned trigonal carbon along the tetragonal c-axis provide high Vickers hardness with a magnitude close to diamond.*

## 1- Introduction and context

Diamond, as natural gem and man-made for applications, is recognized as the hardest material [1]. In last decades large research efforts were devoted to identifying novel allotropes of carbon close to diamond using modern materials research software as USPEX (Universal Structure Predictor: Evolutionary Xtallography) [2]. Regarding carbon, in view of the large number of claimed stoichiometries and structures, a database: SACADA was built regrouping all known carbon allotropes thus helping researchers in their endeavor and letting avoid claiming existing systems as novel allotropes [3].

Another initial pathway that could be used complementarily to modern materials research software lies in crystal chemistry rationale calling for structural 'engineering'. Recently, body-centered tetragonal $C_4$ (Figure 1a) was proposed by us in space group $I\bar{4}m2$ as one of the simplest three-dimensional (3D) carbon networks with the purpose of serving as seed-template for other original chemical compounds [6]. Using TopCryst crystallography package [7] $C_4$ was found with "dia" topology -for diamond- as categorized for several chemical compounds in Reticular Chemistry Structure Resource (RCSR) Database [8].

In this work, starting from $C_4$, we build novel tetragonal pentacarbon $C_5$ through crystal chemistry followed by full geometry relaxations using calculations based on the quantum mechanics density functional theory DFT [9]. Analyzed through TopCryst, the fully geometry relaxed structure of $C_5$ was found to be in an original topology within RCSR Database. The novel allotrope characterized with both $sp^2$ and $sp^3$ carbon hybridizations was found cohesive, and stable both mechanically and dynamically accompanied with large hardness magnitude slightly below diamond. It needs to be mentioned at this point that a mixed carbon hybridization ($sp^2$ and $sp^3$), cubic 'pentadiamond' was reported in 2020 as a novel allotrope with high mechanical properties close to diamond, but the paper was later retracted due to calculational errors on the mechanical properties admitted by the authors [10]. The relevance of mixed carbon hybridization is in the change of the electronic structure of insulating diamond, bringing metallic-like behaviors leading to applications.

After this contextual introduction (Section 1) and next Section 2 reporting the computational framework, the crystal chemistry rationale leading to novel tetragonal $C_5$ is discussed in Section 3. Section 4 is devoted to illustrating qualitatively the charge density projections on the atoms and between them. The mechanical and dynamic stabilities and properties are detailed in sections 5 and 6 respectively. The electronic band structures and density of states are discussed in Section 7 with a concluding statement.

## 2- Computational framework

For the search for the ground state structures of the devised structures, geometry optimizations calculations onto the ground state with minimal energies were performed using DFT-based plane-wave Vienna Ab initio Simulation Package (VASP) [11,12]. For the carbon atomic potential including valence states, the projector augmented wave (PAW) method was applied [12,13]. Treating at the same level the exchange X and the correlation C, the exchange-correlation (XC) effects were considered using a generalized gradient approximation (GGA) [14]. The relaxation of the atoms onto ground state geometry was done applying a conjugate-gradient algorithm [14]. A tetrahedron method [15] with corrections made with Methfessel-Paxton scheme [16] was applied for geometry optimization and energy calculations. A special $k$-point sampling [17] was applied for approximating the reciprocal space Brillouin-zone (BZ) integrals. For good reliability, the optimization of the structural parameters was carried out along with successive self-consistent cycles while increasing k-mesh until the forces on atoms were less than 0.02 eV/Å and the stress components below 0.003 eV/Å$^3$.

Besides the elastic constants calculated to infer the mechanical stabilities and hardness, calculations of phonon dispersion curves were also carried out to verify the dynamic stability of the new carbon allotropes (template $C_3$ and novel $C_5$). For the purpose, the phonon modes were computed considering the harmonic approximation via finite displacements of the atoms around their equilibrium positions to obtain the forces from the summation over the different configurations. The phonon dispersion curves along the direction of the Brillouin zone are subsequently obtained using "Phonopy" interface code based on Python language [18]. Finally, the electronic band structures and density of states were obtained with the full-potential augmented spherical wave ASW method based on DFT using the same GGA scheme [19].

## 3- Crystal chemistry and characteristics of $C_5$

Body center tetragonal $C_4$ structure (Fig. 1a) consists of two distinct carbon sites with 2-atoms occupancy: C1 at the corner and body center positions developing *C4* tetrahedra with C2 positioned at the faces (Table 1a). We highlight that all lattice parameters of the structures in Table 1 result from unconstrained geometry optimizations to the energy ground states. The removal of the body center carbon breaks the body-center symmetry and leads to $C_3$ where tetrahedra are at the 8 corners of the structure forming a two-dimensional like stacking along c-tetragonal direction (Fig. 1b). Such modification leads to symmetry lowering and the space group changes from $I\bar{4}m2$, N°119 down to $P\bar{4}m2$, N°115. The second column of Table 1 shows $C_3$ crystal parameters with the Wyckoff positions. Both types of carbon atoms are labeled C1(tet) and C2(tet) in so far that they produce tetrahedral carbon. Expectedly, the volume and the interatomic distance decrease with the removal of one carbon.

The last lines present the total energy and the atom averaged cohesive energies obtained from deducting the atomic contribution of a single C. From the atom-averaged cohesive energy there is a large destabilization of the diamond-like structure upon removal of central carbon to create hypothetical $C_3$ meant to be used as template to devise novel allotropes. Despite the small magnitude cohesive energy (Table 1) -versus the other allotropes in Table 1-, $C_3$ was found, nevertheless, as mechanically and dynamically stable from the sets of elastic constants and phonons band structures respectively, as detailed in the development of the paper.

Consequently, the following scheme is presented:

$C_3$ "receives"

- one extra carbon to make $C_4$, already investigated [6],
    - C-C pair to make novel $C_5$ (shown in Fig. 1c),
        - one additional carbon atom at cell center, based on $C_5$, to make $C_6$ introducing C-C-C, published as a structure in CCDC [21] for future developments.

The major difference between $C_4$ and $C_5$ is that whereas the central carbon is tetrahedral C(tet) completing the diamond-like edifice as discussed above, the two additional carbon atoms are labeled C(trig), "trig" standing for trigonal carbon shown with white spheres in Fig. 1c. Alike $C_3$, the new allotrope $C_5$ belongs to $P\bar{4}m2$, N°115 space group (no body-center symmetry). The C(tet)-C(tet) interatomic distances are within range of $C_4$ (diamond-like) and smaller magnitudes are observed for $d(C_{trig}-C_{trig}) = 1.47$ Å. Beside C-C single bonds for C(tet)-C(tet), we are in presence of additional double C=C bond-like, thus characterizing pentacarbon $C_5$ with $sp^2/sp^3$ hybridizations. The atom averaged cohesive energy is significantly larger than in $C_3$ with a magnitude closer to $C_4$.

### 4- Charge density projections.

Further qualitative illustration of the different types of hybridizations is obtained from the projections of the charge densities around and between atoms. Figure 2 presents the projections shown with yellow volumes. Upon crossing a crystal plane, charge density slices with red color are shown indicating strong charge localizations.

In $C_4$ (Fig. 2a) the sp$^3$-type hybridization expected for C(tet) is clearly observed especially on central carbon with the yellow volumes taking the shape of a tetrahedron, and $C_4$, alike diamond, is a perfectly covalent chemical system. Upon removal of central carbon producing $C_3$ with only C(tet) -cf. Table 1-, the charge density is modified especially for the carbon atoms pointing towards the empty space as it is exhibited by the larger red area versus $C_4$. Such behavior resembling dangling bonds is in line with the low cohesive $C_3$.

Larger changes are observed in $C_5$ (Fig. 2c) where the charge density (dangling bonds in $C_3$) is skewed towards the pair of C(trig) white spheres. The charge density in no more homogeneously distributed as in covalent $C_4$, and less pronounced red color is observed letting suggest a polar-covalent behavior normally usually found in compounds with different constituents' electronegativities such as boron nitride BN. Therefore, we are presented with a decrease of the covalence from $C_4$ to $C_5$ brought by the introduction of trigonal carbon.

## 5- Mechanical properties from elastic constants

The investigation of mechanical properties was based on the calculations of the elastic properties determined by performing finite distortions of the lattice and deriving the elastic constants from the strain-stress relationship. Most compounds are polycrystalline, and generally considered as randomly oriented single crystalline grains. Consequently, on a large scale, such materials can be considered as statistically isotropic. They are then fully described by bulk (*B*) and shear (*G*) moduli obtained by averaging the single-crystal elastic constants. The method used here is Voigt's [22], based on a uniform strain. The calculated sets of elastic constants are given in Table 2; the elastic constants of $C_4$ [6] are reported for the sake of comparison. While most elastic constants of $C_4$ are larger than in $C_5$, it can be noted that $C_{33}$ magnitude is slightly larger in $C_5$, concomitantly with aligned trigonal C-C along the c-tetragonal axis (cf. Fig. 1c).

The elastic constants of $C_3$ exhibit a relatively large magnitude for in-plane $C_{11}$ and much smaller magnitude for $C_{33}$ relevant to inter-planes, i.e., along tetragonal c-direction. Both largest $C_{ii}$ ($C_{11}$ and $C_{33}$) are smaller than the corresponding values in $C_5$. Indeed, the chemical system describing $C_3$ is more relevant to two-dimensional 2D letting it receive interstitials as schematized in Section 3, and hence leading to 3D $C_4$, $C_5$, $C_6$ etc. All magnitudes of the other $C_{ij}$'s are very small letting expect a soft material. However, it will be shown that $C_3$ is dynamically valid from the calculations of positive phonon frequencies in next section.

All $C_{ij}$ (i=j and i≠j) values are positive and their combinations obey rules pertaining to the mechanical stability of the chemical system.

$C_{ii}$ (i =1, 3, 4, 6) > 0; $C_{11}$ > $C_{12}$, $C_{11}$+ $C_{33}$ − 2$C_{13}$ > 0;

and 2$C_{11}$+ $C_{33}$ + 2$C_{12}$ + 4$C_{13}$ > 0.

The equations providing the bulk $B_V$ and shear $G_V$ moduli are as follows for the tetragonal system [23]:

$$B_{Voigt}^{tetr.} = 1/9\ (2C_{11} + C_{33} + 2C_{12} + 4C_{13});$$

and

$$G_{Voigt}^{tetr.} = 1/15\ (2C_{11} + C_{12} + 2\ C_{33} − 2C_{13} + 6C_{44} + 3C_{66}).$$

$C_4$ has the largest $B_V$ and $G_V$, close to the accepted values for diamond $B_V$ =445 GPa and $G_V$ = 550 GPa [1]. The corresponding $B_V$ and $G_V$ magnitudes in $C_5$ are slightly smaller but they remain high, oppositely to $C_3$ which shows much smaller $B_V$ and $G_V$ magnitudes versus $C_5$. From the calculated $B_V$ and $G_V$ we can predict the Vickers hardness $H_V$, evaluated with the model of Chen et al. based on the bulk and shear moduli obtained from the elastic constants [24]. $H_V$ is formulated as follows:

$H_V$ = 0.92 (G/V)$^{1.137}$ G$^{0.708}$.

The numerical values are given in the last column of Table 1. $H_V(C_4)$ = 97 GPa is a magnitude in agreement with the admitted values of diamond [1]. $H_V(C_5)$ = 81 GPa is found smaller than in $C_4$, but the large magnitude leads to announce $C_5$ as ultra-hard carbon allotrope characterized with mixed carbon hybridizations. Lastly, $H_V(C_3)$ = 15 GPa letting describe it as a soft material. It should be noted here that Chen et al. model is not the only one to calculate Vickers hardness. Using the thermodynamic model relying on the structure topology [25], $H_V(C_5)$ = 89 GPa, a magnitude close to the one obtained above and closer to diamond.

6- **Dynamic properties from the phonons**

Another criterion of stability is obtained from the phonons defined as quanta of vibrations; their energy is quantized through the Planck constant 'h' used in its reduced form ℏ (ℏ = h/2π) giving with the wave number ω the phonons energy: E = ℏω. Besides the novel allotropes $C_3$ and $C_5$ the phonon band structures of $C_4$ [6] are shown for the sake of comparison. Fig. 3 shows the phonon bands. Along the horizontal direction, the bands run along the main lines of the tetragonal Brillouin zone (reciprocal k- space). The vertical direction shows the frequencies given in units of terahertz (THz). Since no negative frequency magnitudes are observed, expectedly for formerly studied $C_4$, but also for present $C_3$ and $C_5$, all structures are considered as dynamically stable. There are 3N-3 optical modes at higher energy than three acoustic modes which start from zero energy (ω = 0) at the Γ point, center of the Brillouin Zone, up to a few Terahertz. The acoustic modes correspond to the lattice rigid translation modes of the crystal (two transverse and one longitudinal). The remaining bands correspond to the acoustic modes and culminating at ω ~ 40 THz in $C_4$ and $C_5$, a magnitude observed for diamond by Raman spectroscopy [26], and only up to 34 in less stable $C_3$ (cf. Table 1). The phonons show the closeness of $C_5$ to $C_4$ regarding dynamic stability.

### 7- Electronic band structures and density of states

Using the crystal parameters in Table 1, the electronic bands structures shown in Figure 4 were obtained using the all-electrons DFT-based augmented spherical method (ASW) [20]. The bands develop along the main directions of the primitive tetragonal Brillouin zones. For diamond-like insulating $C_4$ (Fig. 4a), the energy level along the vertical line is with respect to the top of the valence band (VB), $E_V$. As a specific character of diamond, the band gap of $C_4$ is indirect along $k_z$ between $Γ_{VB}$ and $Z_{CB}$ with a magnitude close to 5 eV. Oppositely, $C_3$ (Fig. 4b) and $C_5$ (Fig. 4c) behave as metals with bands crossing the Fermi level $E_F$. The delocalized π electrons are likely responsible of the electronic conductivity.

Concluding, with novel $C_5$ we are presented material with metallic-like mixed $sp^2$-$sp^3$-like carbon hybridizations with ultra-hard properties slightly below diamond.

**Acknowledgments**: I am grateful to Dr Vladimir Solozhenko Directeur de Recherche at CNRS-Paris for exchanges on the topic regarding 'pentadiamond', and the hardness properties of the $C_5$ allotrope.

**Author statement**: I declare that conceptual, calculations and analyses are all mine and I have no conflict of interest with any other work or colleagues.

**References**

[1] V.V. Brazhkin, V.L. Solozhenko, Myths about new ultrahard phases: Why materials that are significantly superior to diamond in elastic moduli and hardness are impossible. J. Appl. Phys. 125, 130901 (2019).

[2] A.R. Oganov. Crystal structure prediction: reflections on present status and challenges. Faraday Discuss. 211, 643 (2018 ).

[3] R. Hoffmann, A. A. Kabanov, A. A. Golov, D. M. Proserpio. Homo Citans and Carbon Allotropes: For an Ethics of Citation. Angew. Chem. Int. Ed., 55, 10962–10976 (2016). & SACADA database (*Samara Carbon Allotrope Database*). www.sacada.info

[6] S.F. Matar, V.L. Solozhenko. The simplest dense carbon allotrope: Ultra-hard body centered tetragonal $C_4$. J. Solid State Chem. 314, 123424 (2022) & Corrigendum to "The simplest dense carbon allotrope: Ultra-hard body-centered tetragonal $C_4$"
https://doi.org/10.1016/j.jssc.2022.123587

[7] A.P. Shevchenko, A.A. Shabalin, I.Yu. Karpukhin, and V.A. Blatov Topological representations of crystal structures: generation, analysis and implementation in the TopCryst system, Science and Technology of Advanced Materials: Methods, 2:1, 250-265 (2022).

[8] O'Keeffe M., Peskov M. A., Ramsden S. J., Yaghi O. M. The reticular chemistry structure resource (RCSR) database of, and symbols for, crystal nets. Acc. Chem. Res. 41, 1782-1789 (2008).

[9] P. Hohenberg, W. Kohn, Inhomogeneous electron gas. Phys. Rev. B 136, 864-871 (1964); & W. Kohn, L.J. Sham, Self-consistent equations including exchange and correlation effects. Phys. Rev. A 140, 1133-1138 (1965).

[10] Y. Fujii, M. Maruyama, N. Thanh Cuong, S. Okada. Pentadiamond: A Hard Carbon Allotrope of a Pentagonal Network of sp2 and sp3 C Atoms. Phys. Rev. Lett. 125, 016001 (2020) & Retraction: Phys. Rev. Lett. 125, 079901 (2020).

[11] G. Kresse, J. Furthmüller, Efficient iterative schemes for ab initio total-energy calculations using a plane-wave basis set. Phys. Rev. B 54 11169, (1996).

[12] G. Kresse, J. Joubert, From ultrasoft pseudopotentials to the projector augmented wave. Phys. Rev. B 59 1758-1775 (1999).

[13] P.E. Blöchl, Projector augmented wave method. Phys. Rev. B 50 (1994) 17953-17979.

[14] J. Perdew, K. Burke, M. Ernzerhof, The Generalized Gradient Approximation made simple. Phys. Rev. Lett. 77 3865-3868 (1996).

[15] W.H. Press, B.P. Flannery, S.A. Teukolsky, W.T. Vetterling, Numerical Recipes, 2nd ed. Cambridge University Press: New York, USA, 1986.

[16] P.E. Blöchl, O. Jepsen, O.K. Anderson, Improved tetrahedron method for Brillouin-zone integrations. Phys. Rev. B 49 16223-16233 (1994)

[17] H.J. Monkhorst, J.D. Pack, Special k-points for Brillouin Zone integration. Phys. Rev. B 13 (1976) 5188-5192.

[18] A. Togo, I. Tanaka. First principles phonon calculations in materials science", Scr. Mater., 108, 1-5 (2015).

[20] V. Eyert, Basic notions and applications of the augmented spherical wave method. Int. J. Quantum Chem. 77, 1007-1031 (2000).

[21] S.F. Matar, CCDC 2208344: (Refcode NEQHUQ; name C6POLr), 2022.

DOI:10.5517/ccdc.csd.cc2d3yxd

[22] W. Voigt, Über die Beziehung zwischen den beiden Elasticitätsconstanten isotroper Körper. Annal. Phys. 274, 573-587 (1889).

[23] D.C. Wallace, Thermodynamics of crystals. New York, USA: John Wiley and Sons; 1972.

[24] Xing-Qiu Chen, Haiyang Niu, Dianzhong Li, Yiyi Li, Modeling hardness of polycrystalline materials and bulk metallic glasses, Intermetallics 19 1275–1281 (2011).

[25] Mukhanov, V.A., Kurakevych, O.O. & Solozhenko, V.L. The interrelation between hardness and compressibility of substances and their structure and thermodynamic properties. J. Superhard Mater. **30**, 368–378 (2008). https://doi.org/10.3103/S1063457608060026

[26] R.S. Krishnan, Raman spectrum of diamond, Nature 155 171 (1945).

Table 1. Tetragonal carbon crystal structure parameters: $C_4$, $C_3$, and $C_5$. Lattice constants and distances are in units of Å (Volume in Å$^3$). Energies are in eV. Atomic energy of C: -6.6 eV

|  | $C_4$ [6] | $C_3$ | $C_5$ |
|---|---|---|---|
|  | $I\bar{4}m2$, N°119 | $P\bar{4}m2$, N°115 | $P\bar{4}m2$, N°115 |
| $a$ | 2.527 | 2.521 | 2.4786 |
| $c$ | 3.574 | 3.402 | 5.0279 |
| C1(tet.). | (2$a$) 0, 0, 0 | (1$a$) 0, 0, 0 | (1$a$) 0, 0, 0 |
| C2(tet.) | (2$d$) ½, 0, ¼ | (2$g$) ½, 0, $z$  $z$= 0.237 | (2$g$) ½, 0, $z$  $z$= 0.187 |
| C(trig.) | - | - | (2$f$) ½, ½, $z$  $z$= 0.646 |
| Volume | 22.82 | 21.63 | 30.89 |
| $d_{C1(tet.)-C2(tet.)}$ | 1.547 | 1.50 | 1.55 |
| $d_{C2(tet.)-C2(trig.)}$ | - |  | 1.50 |
| $d_{C(trig)-C(trig.)}$ | - | - | 1.46 |
|  |  |  |  |
| $E_{total}$. | -36.36 | -21.58 | -43.26 |
| $E_{coh./at.}$ | -2.49 | -0.59 | -2.05 |

Table 2. Calculated elastic constants and bulk $B_V$ and shear $G_V$ moduli. Values are in units of GPa

|  | $C_{11}$ | $C_{12}$ | $C_{13}$ | $C_{33}$ | $C_{44}$ | $C_{66}$ | $B_{Voigt}$ | $G_{Voigt}$ | $H_{Vickers}$ |
|---|---|---|---|---|---|---|---|---|---|
| $C_4$ [6] | 1147 | 28 | 126 | 1050 | 461 | 559 | 434 | 574 | 97 |
| $C_5$ | 918 | 10 | 122 | 1169 | 197 | 361 | 390 | 414 | 81*  89** |
| $C_3$ | 696 | 20 | 33 | 155 | 10 | 5 | 191 | 115 | 15 |

*Using Chen et al. model [24]

**Using Mukhanov et al. thermodynamic model [25].

FIGURES

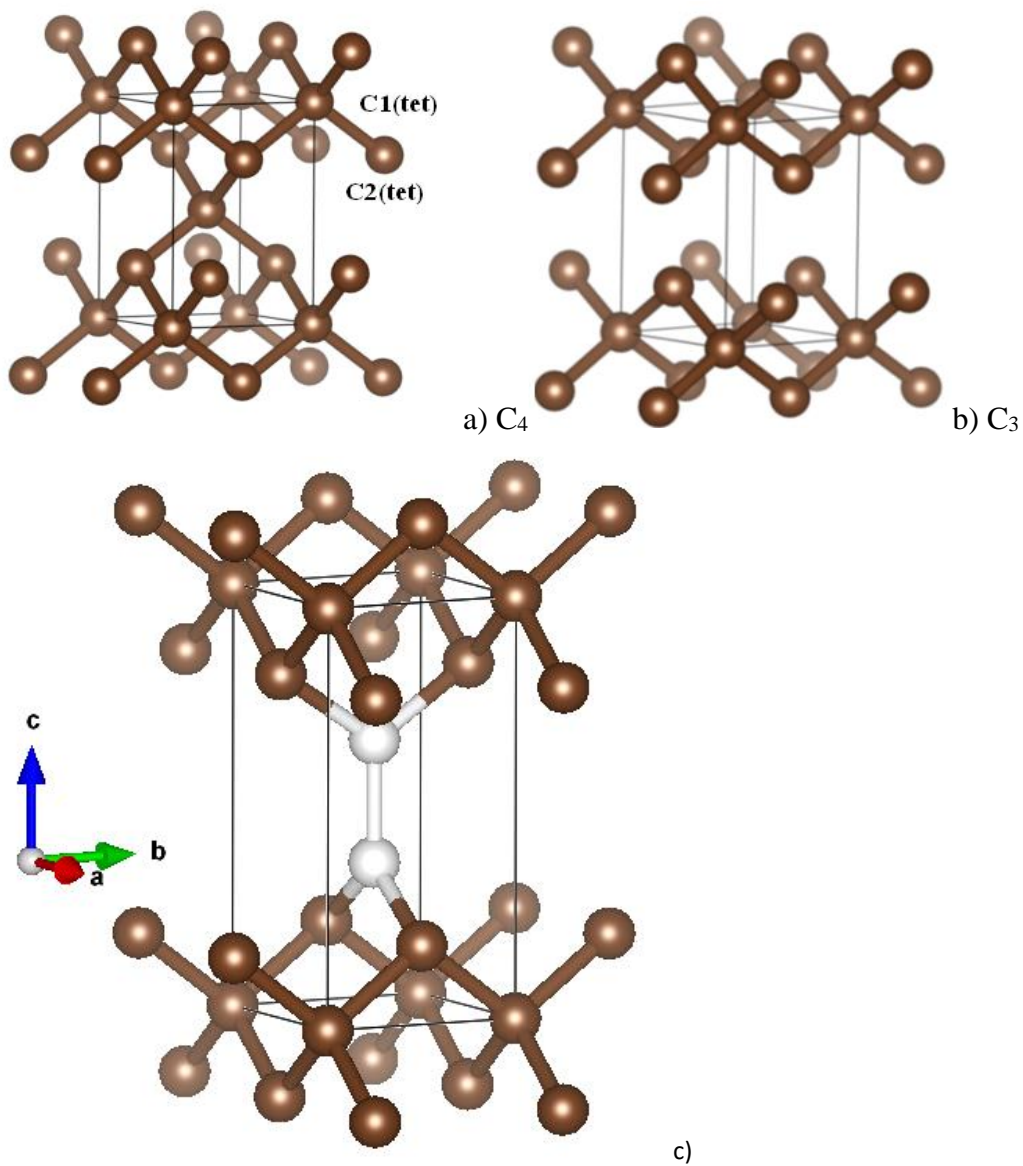

a) C$_4$    b) C$_3$

c)

Figure 1: Sketches of the crystal structures. a) C$_4$ [6], b) model C$_3$, c) novel pentacarbon C$_5$ shown with white spheres depicting trigonal carbon C=C pair.

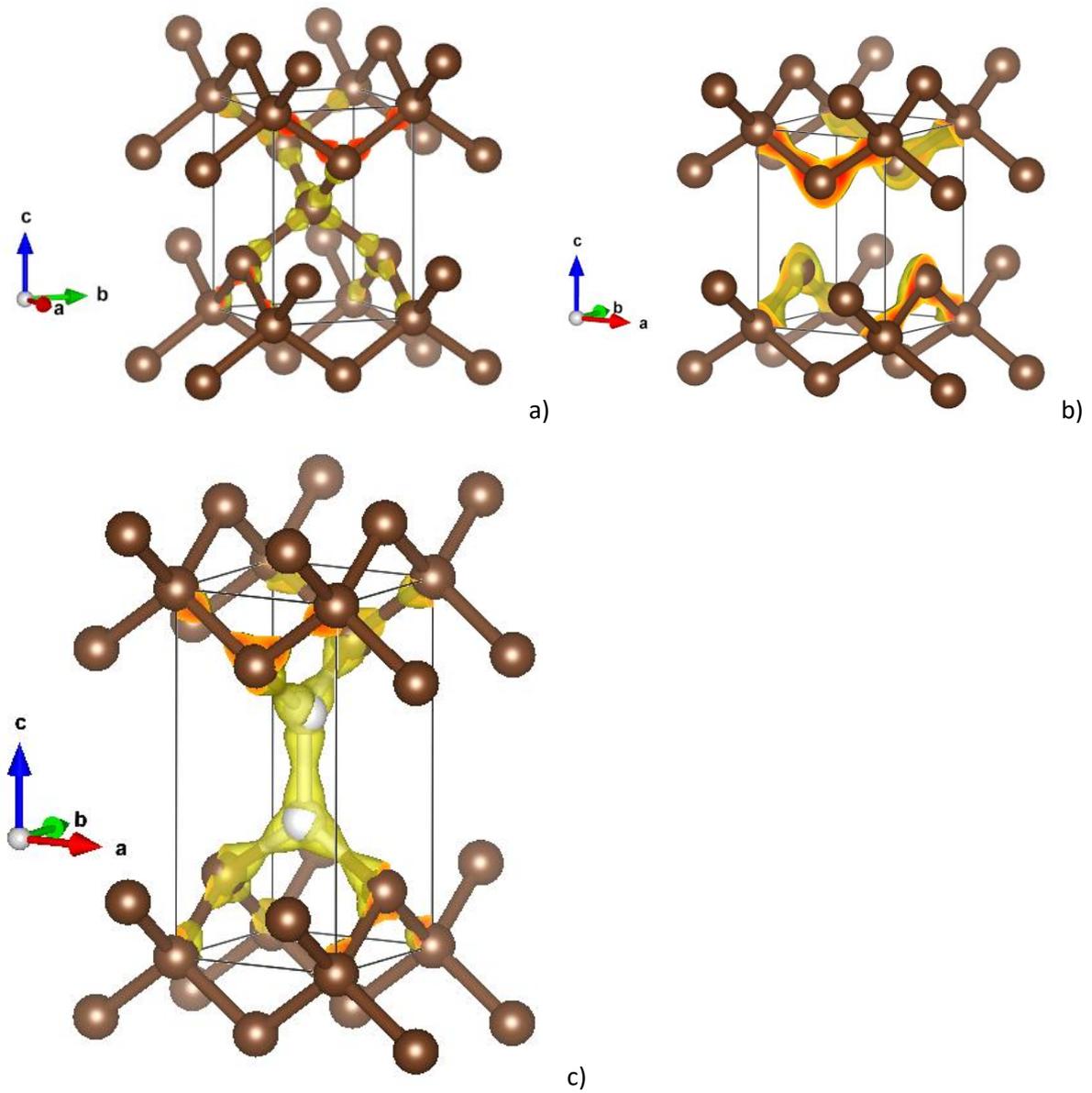

Figure 2: Charge density projections (yellow volumes) in a) $C_4$, b) $C_3$, and c) $C_5$

White spheres represent trigonal C=C carbon pair.

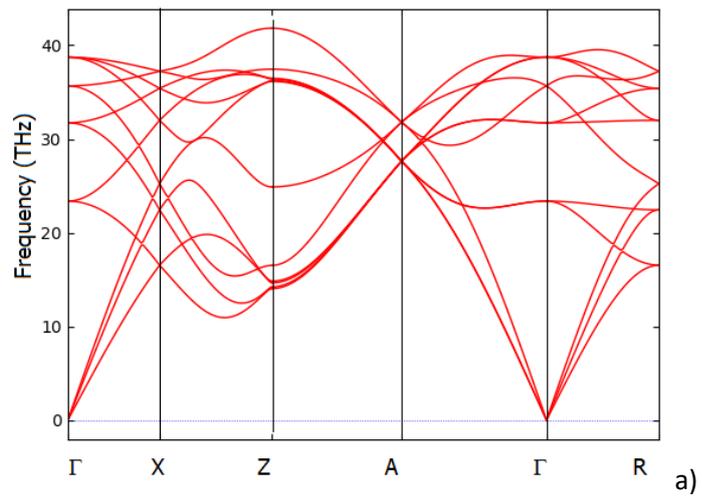

a) C₄

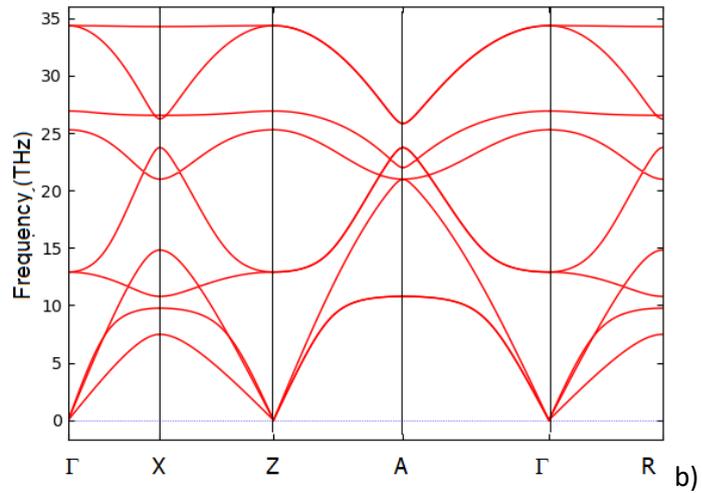

b)

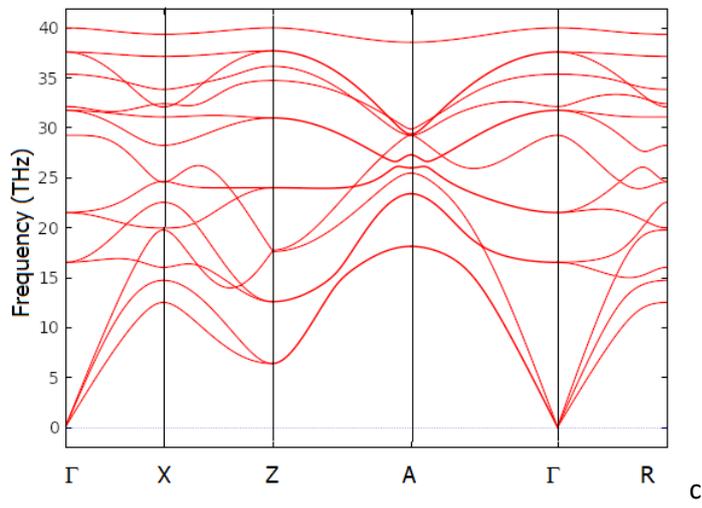

c)

Figure 3: Phonons band structures of a) $C_4$ [6], b) $C_3$, and c) $C_5$

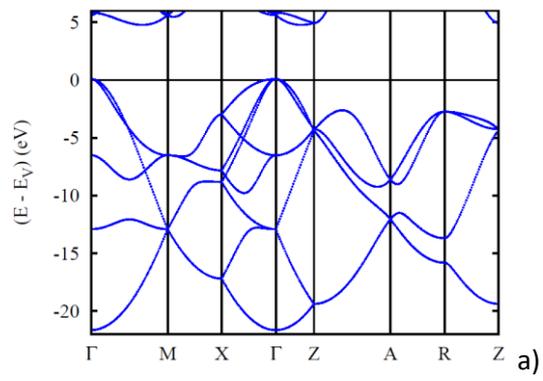

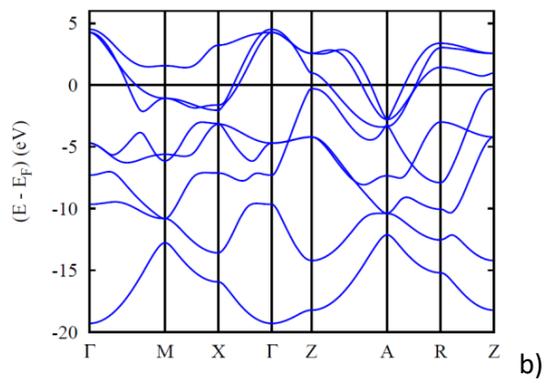

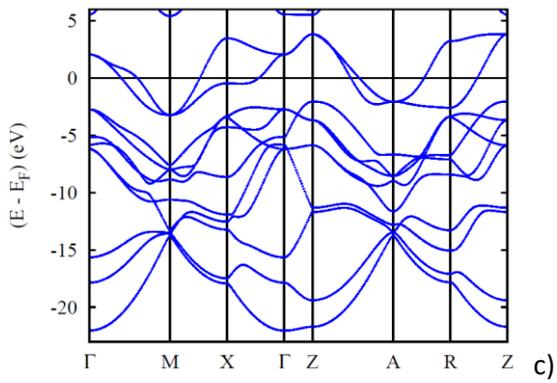

Figure 4: Electronic band structures: a) $C_4$,[6], b) $C_3$, c) $C_5$.